# Effects of Spin Fluctuations and Anomalous Thermal Expansion of delta-Plutonium

A. Solontsov[1,2,3] and V.P. Antropov[1]

[1]*Ames Laboratory USDOE, Ames, IA 50011 USA*
[2]*A.A. Bochvar Institute for Inorganic Materials, 123060 Moscow, Russia*
[3]*State Center for Condensed Matter Physics, M. Zakharova str., 6/3, 155569, Moscow, Russia*

**Abstract.**
We suggest a model for the magnetic dynamics of $\delta$- plutonium and its alloys in order to show that the dynamical fluctuations of the magnetization density, or spin fluctuations, may be responsible for the anomalies of their observed thermal expansion. We show that due to strong magneto-elastic coupling, spin fluctuations may essentially contribute to the volume strain by giving a negative magneto-volume contribution that is proportional to the squared local magnetic moment and the magnetic Gruneisen constant which is negative in $\delta$- plutonium. In the presented model, the local magnetic moment increases as the temperature rises, resulting in the interplay between the positive contributions to the volume strain from the lattice and the negative contribution from spin fluctuations, and finally leads to the Invar anomaly or to the negative coefficient of thermal expansion. Our results agree closely with the measured thermal expansion data for Pu-Ga alloys.

After decades of extensive investigation, plutonium and its alloys still contain many puzzling points for researchers. Structurally, plutonium possesses a cascade of phase transformations resulting in a 25% volume collapse between $\alpha-$ and $\delta-$phases. This change in volume is crucial for the problems of plutonium's technical usage. Though these phase transitions are well known, their mechanisms are far from being understood. From a practical point of view, the $\delta-$phase is most interesting because it is stable above 593K and may be stabilized at lower temperatures by adding a small amount of *Ga* and *Al,* for example (see the review [1]). The mechanism of stabilization of the $\delta-$phase remains unclear. The $\delta-$phase of *Pu* and its alloys possesses a lot of anomalous properties among which the most important is its negative thermal expansion which turns to exhibit an Invar anomaly or become positive on increasing the concentration of stabilizing elements.

Possible ordered magnetism of $\delta-Pu$ would certainly explain its anomalous thermal expansion [2,3]. However, in spite of numerous band-structure calculations based on various applications of the density functional theory, starting with the first relativistic approach [4] (for review see, for instance, Ref. [1]), which predicted ferro- or antiferromagnetic order, no ordered magnetism was observed in $\delta-Pu$ and its alloys [1].

Since then, many electronic structure studies were performed and they offer a large variety of explanations of the absence of magnetism in $\delta-Pu$ (see, e.g., the recent papers [5-7] and references therein). However, the overall picture following from the band-structure studies is still close to suggested by Johansson [8]: $\delta-Pu$ may be viewed as an itinerant system with a borderlike itinerant vs localized f-electron behaviour. These studies point to the itinerant character of 5f-electrons in *Pu* which result in a peak of the density of electronic states at the Fermi level supported by photoemission measurements [9]. In addition, this agrees with the measurements of the specific heat, which, for the Sommerfeld coefficient characterizing the electronic contribution in the stabilized $\delta-Pu,$ is about 65 mJK$^{-2}$mol$^{-1}$ (see [1]), and is close to those in systems with heavy fermions.

Below we shall concentrate on magnetic properties of $\delta$-*Pu*. Its magnetic susceptibility is rather high and is comparable to the low-temperature susceptibility of the itinerant antiferromagnet *Mn* [1], and this points to nearly magnetic behaviour of $\delta$-*Pu* where an essential role is played by the dynamical fluctuations of magnetization density of the electron Fermi-liquid, or spin fluctuations (SF). Effects of SF in *Pu* are widely discussed since 70-s with respect to its unusual transport properties [10]. Due to strong magneto-elastic coupling in *Pu* that manifests in the discontinuities of the temperature dependence of the magnetic susceptibility at structural transitions (see Fig.2 in [1]), SF should be strongly coupled to the crystal lattice, and may essentially affect thermal expansion, phonon spectra, structural phase transitions, and phase stability of different phases of *Pu*. In the present paper, we argue that an account of SF effects in *Pu* may be the clue needed for understanding its anomalous thermal expansion.

The first indications of effects of spin fluctuations in *Pu* were discovered in 1972, when Arko et.al. [11] showed that the $\sim T^2$ dependence of the low-temperature electroconductivity in $\alpha-Pu$ should be related to the scattering of conductive electrons by SF induced by f-electrons. Since then, these results were extended to the $\delta-$ phases of *Pu* stabilized by *Al, Ga, Am*, etc. [12] (For an earlier work, see [13]).

Other evidence for SF in *Pu* was demonstrated by nuclear magnetic resonance (NMR) measurements in the stabilized $\delta-$ phase of *Pu-Ga* alloys [14] at temperatures below 650K.

A recent attempt to analyse SF in the $\delta-$ phase of the *Pu-Ga* alloy using an inelastic neutron scattering technique [1], which is the most direct method, failed to present evidence of their existence. One of the reasons for this failure may be related to the relatively small range of the energies ($\sim 20 meV$) used in the experiment. This range was bases on the simple estimate of the characteristic SF energy $\sim 5 meV$ taken from the Sommerfeld constant [1] which is probably an unreliable value. One should expect higher SF energies in *Pu*, i.e., similar to the energies of $\sim 100 meV$ observed in Ce [15] which is often treated as a rare-earth analog of plutonium. In the present paper from the analysis of the thermal expansion anomalies of the phase of the *Pu-Ga* we infer the SF energy about 187*meV,* which is only 2 times smaller than that observed by inelastic neutron scattering in $\alpha$ - Ce. According to the authors of Ref.1 further neutron experiments on *Pu* are planned, which can contribute to our understanding of magnetic dynamics of *Pu*. In view of observation of the effects of SF in $\delta-Pu$ by other means we believe that the inelastic neutron measurements of SF in plutonium is a question of time.

Properties of SF are tightly related to the temperature dependence of the static magnetic susceptibility, which has been measured for *Pu* by different methods [1] since the 1960's. Fig.2 in Ref.[1] shows the results of a later analysis of the temperature dependence of the magnetic susceptibility of different phases of *Pu* together with the data [16] for the $\delta-$ phase alloy $Pu_{0.94}Ga_{0.06}$ and antiferromagnetic Mn, and from this, one can make several conclusions. First, the change of the susceptibility of *Pu* in the temperature interval 4-913 K, and up to the melting point, is less than 15%, which is small. Second, the susceptibility of *Pu* experiences discontinuities at the temperatures of structural phase transitions $\alpha \to \beta \to \gamma \to \delta \to \delta' \to \varepsilon$, and is accompanied by jumps in volume and the lattice parameter at the phase transition temperatures with an expansion of about 25% between $\alpha$- and $\delta$-phases (as shown in Fig.1). This firmly suggests a strong magneto-elastic coupling giving rise to the interaction of spin modes with the crystal lattice. The values of low-temperature susceptibilities of $\alpha$- and $\delta$-Pu are close to the values of the susceptibilities of *Mn* and *Pd* [1,17], respectively, one being an itinerant antiferromagnet and the other an exchange-enhanced paramagnet. This is another argument to consider $\delta$-Pu as an itinerant paramagnet where strongly correlated 5f-electrons form narrow bands of quasiparticles of nearly magnetic Fermi-liquid. Importantly, the temperature dependence of the magnetic susceptibility of *Pu* shows no signs of the Curie-Weiss behavior up to the melting point [1], though there are attempts to present it in terms of the phenomenological "modified" Curie-Weiss law [16].

As shown in Fig.1, thermal expansion of $\delta-$Pu has anomalies which are described by the negative coefficient of thermal expansion. In $\delta-$phase *Pu* alloys, and with concentrations of the doping elements *Ga* and *Al* are less than $2at\%$, the coefficient of thermal expansion is either still negative or exhibits the Invar effect [12,18] (see Fig.2). When concentrations of these doping elements are above $3at.\%$, the coefficient of thermal expansion becomes positive (see Fig.3), and until the doping element's concentration rises to $\sim(6-7)at\%$, it remains anomalously small [12,18].

Figs.2 and 3 show the results of neutron diffraction measurements of thermal expansion in $\delta-$phase alloys $Pu_{0,98}Ga_{0,02}$ and $Pu_{0,96}Ga_{0,04}$ in the temperature range 15-800K [18]. In these figures, we also present the calculated lattice contribution to thermal expansion of these alloys within the Debye model, for which we have used the following parameters for *Pu* [18]: $\Theta_D = 131K - 0,047T$, $\Gamma_D = 0.5$, and $B \approx 0,289 \cdot 10^{12} dyn\, cm^{-2}$ for the Debye temperature, Gruneisen constant, and bulk modulus, respectively. As it follows from these figures, thermal expansion of $\delta-$phase alloys of *Pu* is less than was predicted by the Debye model. It should be emphasized that at temperatures of ~600K, the negative non-lattice and lattice contributions to the volume change are comparable, and for alloys with $2at.\%Ga$ and $4at.\%Ga$ they approximately consist 50% and 30%, respectively, of the values predicted by the Debye model. In Fig.4, we present the non-lattice contribution to thermal expansion for $Pu_{0,96}Ga_{0,04}$. We describe it as the difference between the experimental data and calculated results within the Debye model. As shown in Fig.4, the non-lattice thermal expansion changes non-linearly with temperature in the range 15-800K. This shows that the possible excitations responsible for the negative expansion must have energies higher than the energy used in the inelastic neutron scattering experiment (~20 meV) [1].

Let us briefly discuss the possible mechanisms of non-lattice contributions to thermal expansion of $\delta-$Pu. The conventional electronic contribution to the volume change due to Fermi quasiparticles depends quadratically on the temperature ($\sim \Gamma_e (k_B T/\varepsilon_F)^2$, where $\varepsilon_F$ is the Fermi energy, and is proportional to the Gruneisen constant of electrons $\Gamma_e$, which is essentially positive in metals. This strongly suggests that the anomalies of thermal expansion of $\delta - Pu$ should be related to the Bose excitations of the electronic Fermi liquid.

Charge density fluctuations can also contribute to thermal expansion of metals [19]. This contribution is analogous to the electronic one and is proportional to the Gruneisen constant of longitudinal phonons $\sim \Gamma_D$, which is also positive in *Pu*. Thus, the contributions to thermal expansion of $\delta - Pu$ due to electrons and charge density fluctuations are positive and cannot account for the negative expansion shown in Fig.4.

The importance of fluctuations in the electronic system of plutonium and its alloys was recently demonstrated within a simple "Invar" model for $\delta$ – phase plutonium [18]. Actually, they were described as an ensemble of undamped uncoupled oscillators of a non-magnetic origin with the energy $\Delta E / k_B = 1400 K$ separating energies of two hypothetical atomic states of Pu atoms with different atomic volumes. This "Invar" model for *Pu* is actually related to the old Einstein model of 1907 for a crystal (see Eq.(5) in [18]).

Though the physical basis for the high frequency oscillations in the "Invar" model for metallic *Pu* is not clear, and they were not observed by inelastic neutron scattering [18], the "Invar" model was surprisingly successful in the qualitative description of the anomalies of thermal expansion and bulk modulus in the $\delta$ – phase of *Pu* and *Pu-Ga* alloys. This clearly shows the importance of some "electronic" Bose excitations (or fluctuations of the electronic Fermi liquid) interacting with the crystal lattice of $\delta - Pu$ via a negative Gruneisen constant.

In the present paper, we argue that elastic properties of plutonium may be strongly affected by overdamped magnetic excitations of itinerant electrons, SF, which interactions with the crystal lattice may be described by a negative magnetic Gruneisen constant. We argue that the most likely mechanism of negative thermal expansion of $\delta - Pu$ and its alloys is the result of a magneto-volume effect induced by paramagnetic SF giving the negative contribution to thermal expansion which dominates over the lattice contribution.

The well-known result for the magnetic contribution to the volume strain in the SF theory of itinerant magnets [3] reads as

$$(\frac{\Delta V}{V})_m = \frac{C_m}{B} M_L^2(T), \qquad (1)$$

where $(\Delta V / V)_m$ is the relative change in volume $V$ due to the effects of SF and

$$C_m = -\frac{1}{2}\frac{\partial \chi^{-1}}{\partial \ln V} = \frac{1}{2}\chi^{-1}\Gamma_m, \qquad (2)$$

is the conventional magneto-elastic parameter [2] related to the magnetic Gruneisen constant $\Gamma_m$, $\chi = \chi(V,T)$ is the static magnetic susceptibility.

Here, $M_L^2(T)$ is the squared local magnetic moment, or squared averaged amplitude of SF, being the key parameter in the SF theory of itinerant electron magnets [20], and is given by the fluctuation-dissipation theorem [21]

$$M_L^2 = 12\hbar \sum_{\mathbf{Q},\omega} \text{Im} \chi(\mathbf{Q},\omega)\left(N_\omega + \frac{1}{2}\right) \equiv \left(M_L^2\right)_{Z.P.} + \left(M_L^2\right)_T, \qquad (3)$$

where $\chi(\mathbf{Q},\omega)$ is the dynamical magnetic susceptibility dependent on the wavevector $\mathbf{Q}$ and frequency $\omega$ of SF, $\sum_\omega = \int_0^\infty d\omega/2\pi$. The factors $N_\omega = [\exp(\hbar\omega/k_B T)-1]^{-1}$ and 1/2 are related to the thermal $\left(M_L^2\right)_T$ and zero-point $\left(M_L^2\right)_{Z.P.}$ contributions to the squared local moment (3), respectively.

The main temperature dependence to the volume strain (Eq.(1)) results from the temperature variation of the squared local magnetic moment (Eq.(3)). The coupling parameter $C_m$ (Eq.(2)) is weakly temperature dependent due to the weak temperature variation of the magnetic susceptibility in $\delta$-Pu. Because of the lack of experimental data on Pu-Ga alloys, we shall assume that the magnetic Gruneisen constant $\Gamma_m$ and bulk modulus $B$ are temperature independent.

The imaginary part of the dynamical susceptibility in Eq.(3) describes the spectrum of paramagnetic SF in $\delta$-Pu. At the moment, no reliable data on the spatial dispersion of the dynamical magnetic susceptibility of Pu is available, so we shall neglect its wavevector dependence which for not very high frequencies results in [15]

$$\text{Im}\,\chi(\mathbf{Q},\omega) \approx \text{Im}\,\chi(\omega) = \chi\frac{\omega\omega_{SF}}{\omega_{SF}^2 + \omega^2}, \qquad (4)$$

where $\omega_{SF}$ is the characteristic frequency of SF that we treat as an empirical parameter to be inferred from experiments. It should be noted that Eq.(4) describes quasielastic SF localized on Pu atoms that give rise to dynamical magnetic moments with the direction and amplitude varying with the characteristic frequency $\omega_{SF}$. Their averaged squared values are given by Eq.(3). It should be noted that localized SF described by Eq.(4) were observed by inelastic neutron scattering in Ce, an analogue of Pu [15].

The expression for magnetic susceptibility (Eq.(4)) allows us to obtain the following explicit expressions for the zero-point and thermal contributions to the local magnetic moment, respectively, from Eq.(3):

$$(M_L^2)_{Z.P.} = \frac{3v\hbar\omega_{SF}\chi}{2\pi\Omega}\ln[1+(\frac{\omega_c}{\omega_{SF}})^2], \qquad (5)$$

$$(M_L^2)_T = \frac{3k_B T}{\Omega}v\chi g(\frac{\hbar\omega_{SF}}{2\pi k_B T}), \qquad (6)$$

where $v$ is the number of Pu atoms in the unit cell (for fcc $\delta-Pu$ $v=4$), $\Omega$ is its volume, $g(y) = 2y[\ln y - 1/2y - \Psi(y)]$, and $\Psi(y)$ is the Euler's psi function.

6Here, we use Eqs.(1), (2), and (6) to calculate the magnetic contribution to the thermal expansion of the $\delta$ – phase alloy $Pu_{0,96}Ga_{0,04}$ with account of the temperature dependent magnetic susceptibility $\chi$ for $Pu_{0,94}Ga_{0,06}$, as measured in [16]. Fig.4 shows the fit of the calculated results to the non-lattice contribution taken from Fig.3 and gives the value for the temperature independent SF energy $\hbar\omega_{SF} \approx 0,187 eV$, and the magnetic Gruneisen constant $\Gamma_m \approx -1,75$. As it follows from Fig.4, the magnetic contribution to the thermal expansion calculated within a simple model for SF (Eq.(4)) closely resembles the measured thermal expansion data for *Pu-Ga* alloys.

It should be mentioned that for simplicity, we neglect the positive contributions to the thermal expansion arising from the electronic Fermi excitations and charge fluctuations throughout this study. Their account would result in an increase of the non-lattice contribution shown in Fig.4, and would roughly lead to increase in the absolute value of the magnetic Gruneisen constant.

It is necessary to note that our estimate of the characteristic SF energy $\sim 187 meV$, and is only 2 times less than the SF energy measured [15] in $\alpha - Ce$ with the help of inelastic neutron scattering. It is also close to the excitation energy ($\Delta E \sim 120 meV$) estimated [18] using the "Invar" model for *Pu-Ga* alloys. Our estimate for the magnetic Gruneisen constant (-1.75) is comparable with the value (-3) found [18] in the "Invar" model for $Pu_{0,96}Ga_{0,04}$.

In summary, we use a model of spin fluctuations to illustrate a magnetic mechanism of negative thermal expansion in $\delta - Pu$ and its alloys caused by strong coupling of SF to the crystal lattice. This results in an anomalous negative magneto-volume effect dominating over the lattice contribution to the thermal expansion. The used here simple model of magnetic dynamics of plutonium, surely, does not present a quantitative description of its anomalous behavior, but rather point to the necessity to study effects of spin fluctuations, which may play an important role in plutonium.

A.S. is pleased to thank V.K. Orlov and S.A. Kiselev for fruitful discussions and A.A. Burmistrov for the technical support. Work at the Ames Laboratory was supported by Department of Energy-Basic Energy Sciences, under Contract No. DE-AC02-07CH11358. A.S. would also acknowledge the support of ROSATOM and Russian Foundation for Basic Research (grant No 06-02-17291).

**Literature.**

**Figure captions.**

Fig.1. Change of the lattice parameter of *Pu* and *Pu − Ga* alloys at the phase transitions.

Fig.2. Temperature expansion of $Pu_{0.98}Ga_{0.02}$: full circles are results of neutron diffraction measurements [18], full squares are the calculated Debye contribution to the volume change.

Fig.3. Temperature expansion of $Pu_{0.96}Ga_{0.04}$: full circles are results of neutron diffraction measurements [18], full squares are the calculated Debye contribution to the volume change.

Fig.4. Non-lattice contribution to the volume change of $Pu_{0.96}Ga_{0.04}$: triangles are the results of neutron diffraction measurements [18] after the subtraction of the Debye contribution, the curve shows the results of the calculation of the magnetic contribution from Eq.(1), (2) and (6) with the SF energy $\hbar\omega_{SF} \approx 0.187 meV$ and the magnetic Gruneisen constant $\Gamma_m \approx -1,75$.

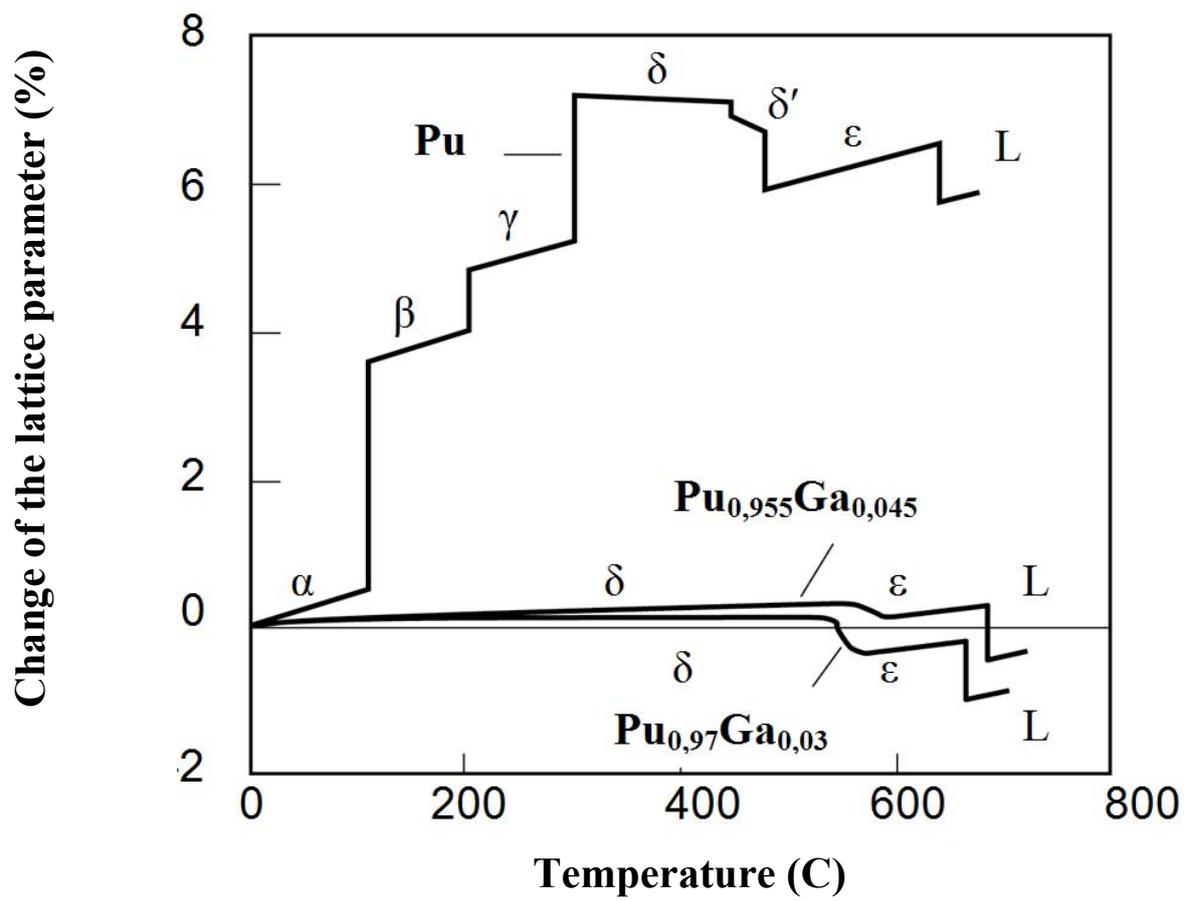

Fig.1

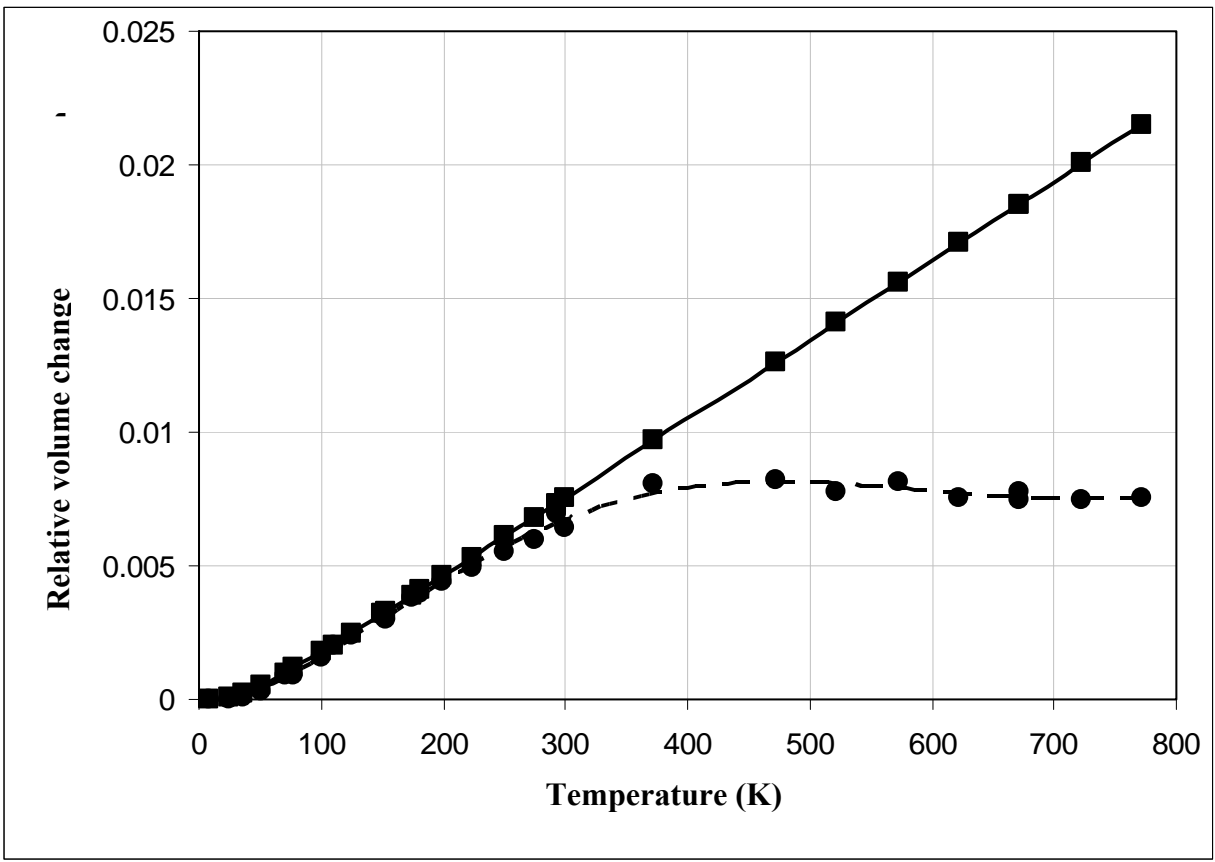

Fig.2

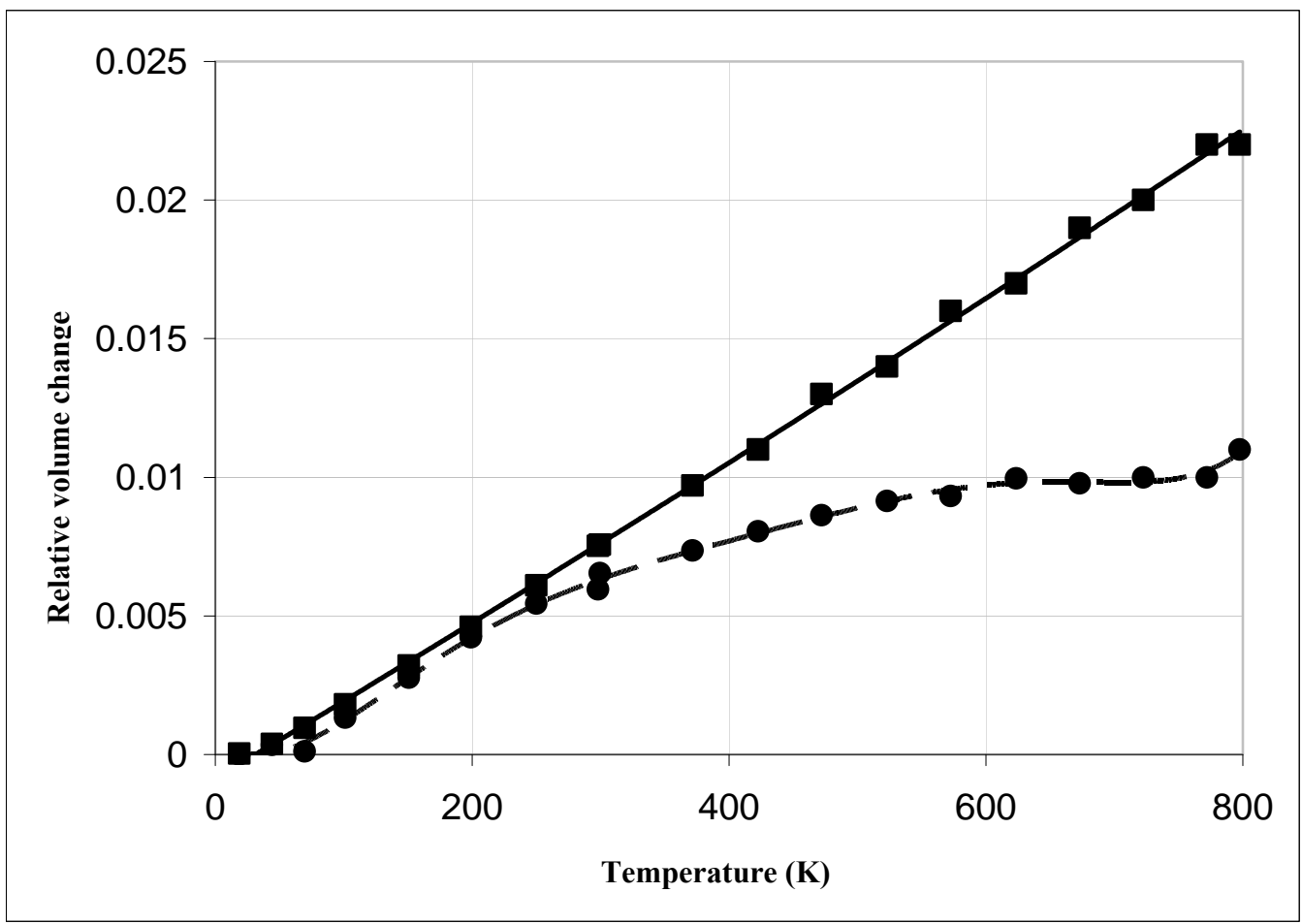

Fig.3

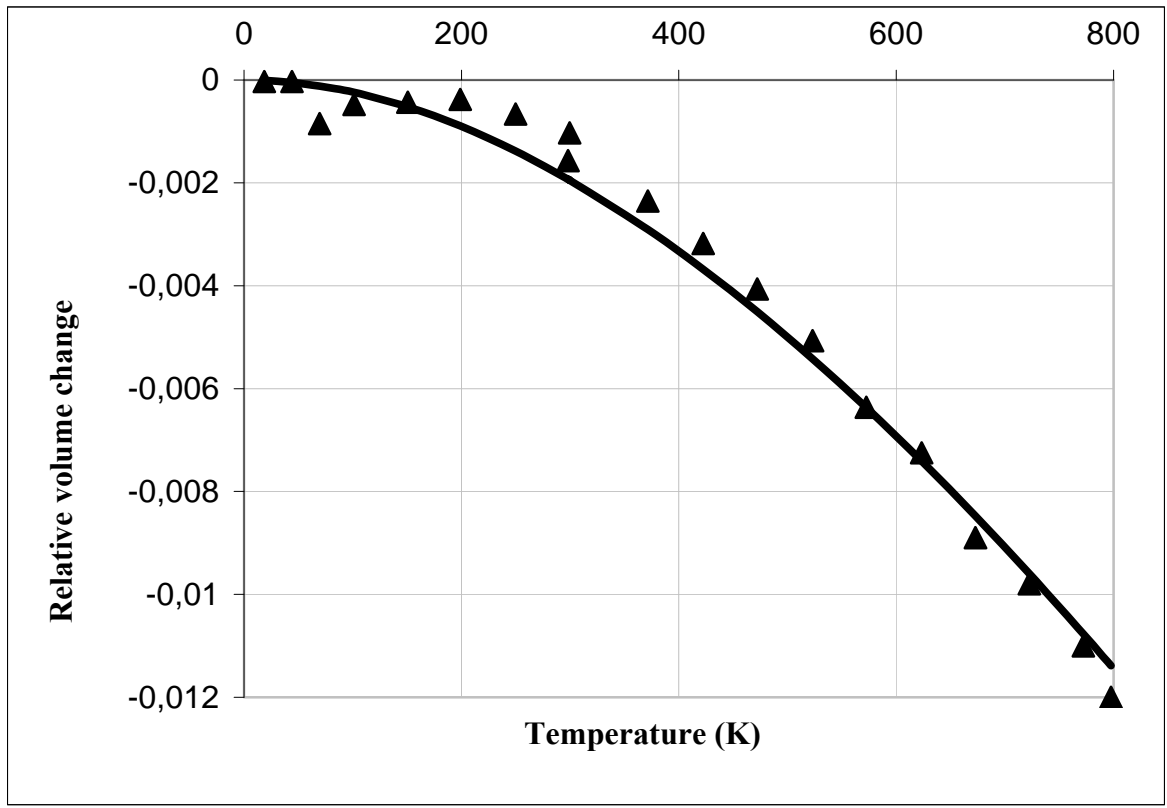

Fig.4